\documentclass[twocolumn,prl]{revtex4-1}

 \usepackage{amssymb}
 \usepackage{dsfont}
 \usepackage{amsmath}
\usepackage{graphicx}
\usepackage{dcolumn}
\usepackage{bm}

\begin{document}

\title{X-matrices provide a lower bound of concurrence}
\author{S. M. Hashemi Rafsanjani and  S. Agarwal}
\affiliation{ Rochester Theory Center and the Department of Physics~\& Astronomy\\
University of Rochester, Rochester, New York 14627}
\email{hashemi@pas.rochester.edu}


\date{\today}

\begin{abstract}
By focusing on the X-matrix part of a density matrix of two qubits we provide an algebraic lower bound for the concurrence. The lower bound is generalized for cases beyond two qubits and can serve as a sufficient condition for non-separability for bipartite density matrices of arbitrary dimension. Experimentally, our lower bound can be used to confirm non-separability without performing a complete state tomography.\end{abstract}

\pacs{..........}

\maketitle

 \newcommand{\beq}{\begin{equation}}
 \newcommand{\eeq}{\end{equation}}
 \newcommand{\bel}{\begin{align*}}
 \newcommand{\tamam}{\end{align*}}
 \newcommand{\dg}[1]{#1^{\dagger}}
 \newcommand{\reci}[1]{\frac{1}{#1}}
 \newcommand{\ket}[1]{|#1\rangle}
 \newcommand{\nim}{\frac{1}{2}}
 \newcommand{\om}{\omega}
 \newcommand{\te}{\theta}
 \newcommand{\la}{\lambda}
 \newcommand{\beqa}{\begin{eqnarray}}             
 \newcommand{\eeqa}{\end{eqnarray}}               
 \newcommand{\nn}{\nonumber}                      
 \newcommand{\bra}[1]{\langle#1\vert}                 
 \newcommand{\ipr}[2]{\left\langle#1|#2\right\rangle}
  \newcommand{\up}{\uparrow}
   \newcommand{\down}{\downarrow}
     \newcommand{\dn}{\downarrow}         

Entanglement, both as a fundamental consequence of quantum mechanics and as a resource that can be utilized in quantum computation, has remained an important subject of investigation in the last two decades \cite{nielsen:2000,horodecki2009} and questions regarding quantifying entanglement of a given state 
continue to challenge physicists. These are questions that are needed to be answered before meeting the challenges regarding the dynamical properties and/ or questions about quantifying the entanglement experimentally. 

By introducing the concurrence \cite{wootters}, Wootters effectively solved the mathematical problem of quantifying the entanglement of a bipartite state shared by two qubits (mixed or pure). He has provided an analytical prescription to find the concurrence which can serve as a measure of entanglement. However, although the prescription makes it possible to calculate the concurrence of any bipartite state of two qubits numerically, the diagonalization procedure involved has proved to limit the feasibility of analytical investigations \cite{[{Especially true if all the elements of the density matrix are non-zero; see, e.g.,~}] chen}.

For a special class of density matrices (X-matrices) \cite{yu-2007}, concurrence takes a very simple form. Thus, it will be quite helpful in drawing analytical insights, if we can find a basis in which the density matrix takes the X form \cite{yu-2007,PhysRevLett.93.140404,*yonac1,*yonac2007,*Cui200744,*Zhang2007274,*PhysRevA.77.054301,*PhysRevA.77.012117}. It is always possible to write a density matrix of two qubits, $\hat{Q}$, as a sum of an X density matrix and a remaining matrix, that we call an O-matrix. Assume that the density matrix $\hat{Q}$, given in a product basis $\{\ket{\up,\up},\ket{\up,\dn},\ket{\dn,\up},\ket{\dn,\dn}\}$, reads:
\begin{align}
\hat{Q}=\left(\begin{array}{cccc}Q_{11} & Q_{12} & Q_{13} & Q_{14} \\Q_{21} & Q_{22} & Q_{23} & Q_{24} \\Q_{31} & Q_{32} & Q_{33} & Q_{34} \\Q_{41} & Q_{42} & Q_{43} & Q_{44} \end{array}\right).
\end{align}
Then $\hat{Q}=\hat{X}+\hat{O}$
\begin{align} 
=\left(\begin{array}{cccc}Q_{11} & 0 & 0 & Q_{14} \\0 & Q_{22} & Q_{23} & 0 \\0 & Q_{32} & Q_{33} & 0 \\Q_{41} & 0 & 0 & Q_{44} \end{array}\right)+\left(\begin{array}{cccc}0 & Q_{12} & Q_{13} & 0 \\Q_{21} & 0 & 0 & Q_{24} \\Q_{31} & 0 & 0 & Q_{34} \\0 & Q_{42} & Q_{43} & 0 \end{array}\right).
\end{align}
The X matrix can be thought of as a density matrix of two qubits itself, and so one can associate a concurrence to it. One wonders, is there a relation between the concurrence of the original matrix and the concurrence of the X matrix?

In this paper we will show that the concurrence of the X matrix is always smaller or equal to the concurrence of $\hat{Q}$. This result provides a lower bound for concurrence that is analytically easy to calculate even when the density matrix is not in the X form. We will then generalize our result to the cases beyond two-qubit density matrices and present a sufficient criterion for non-separability of a bipartite mixed state and a lower bound for the concurrence of such states. We will also compare our lower bound with the exact value of concurrence for a class of mixed states for which the exact value of concurrence is known, namely the class of \textit{isotropic} states.

Furthermore, a potential experimental application exists for this otherwise purely algebraic result. Our lower bound requires, in principle, only three density matrix  elements. Thus it can be a useful tool to confirm entanglement of a bipartite system without performing a complete state tomography.\\

 The $\hat{X}$ matrix depends on the basis used for the original density matrix, $\hat{Q}$. The concurrence of $\hat{X}$ takes a very simple form:
\begin{align}\nn
C(\hat{X})&=2~ \text{Max} \{0,|Q_{14}|-\sqrt{Q_{22}Q_{33}},|Q_{23}|-\sqrt{Q_{11}Q_{44}}\}\\
&= \text{Max} \{0,C_{1}(\hat{X}),C_{2}(\hat{X})\}.
\end{align}
In the following we will prove that $C_{1}(\hat{X})\le C(\hat{Q})$. The proof that $C_{2}(\hat{X})\le C(\hat{Q})$ is identical. Note that $C(\hat{X})$ is a basis-dependent quantity and our claim is that it is always smaller than or equal to the concurrence $C(\hat{Q})$ which is a basis-independent quantity.

Let us first assume that $\hat{Q}$ is a pure state, i.e. $\hat{Q}=\ket{\psi}\bra{\psi}$ where:
 \begin{align}
\ket{\psi}=\alpha\ket{\up,\up}+\beta \ket{\up,\dn}+\gamma \ket{\dn,\up}+\delta\ket{\dn,\dn}.
\end{align}
The concurrence of this state is $C(\ket{\psi}\bra{\psi})=2~|\alpha\delta-\beta\gamma|$, and 
 \begin{align}
C_{1}(\ket{\psi}\bra{\psi})=2~|\alpha\delta|-2~|\beta\gamma|.
\end{align}
Using the triangle inequality returns the desired inequality
 \begin{align}C(\ket{\psi}\bra{\psi})\ge |C_{1}(\ket{\psi}\bra{\psi})|.\label{eqnpure}\end{align}

We now turn our attention to the case when $\hat{Q}$ is allowed to be a mixed state. The concurrence of a mixed state is defined as 
\begin{align} C(\hat{Q})=\text{Min} \sum_{i} p_{i}~C(\ket{\psi_{i}}\bra{\psi_{i}}) \label{eqnMixedcon}, \end{align}
where the minimum is taken over all the pure decompositions of $\hat{Q}$:
\begin{align} \hat{Q}= \sum_{i} p_{i}~\ket{\psi_{i}}\bra{\psi_{i}}. \end{align}

For any density matrix of two qubits, it has been shown \cite{wootters,Wootters:2001:EFC:2011326.2011329,*PhysRevA.64.052304} that there is a decomposition of $\hat{Q}$ that minimizes Eq.(\ref{eqnMixedcon}) with a set of pure states, all having the same concurrence. Let us assume this decomposition reads
\begin{align} \hat{Q}= &\sum_{i} P_{i}~\ket{\phi_{i}}\bra{\phi_{i}},\\
C(\hat{Q})=&\sum_{i}P_{i}~C(\hat{Q}^{(i)}),\end{align}
where $\hat{Q}^{(i)}=\ket{\phi_{i}}\bra{\phi_{i}}$. Since $\ket{\phi_{i}}$ 's are pure states, we can use Eq. (\ref{eqnpure}) and write
\begin{align} 
C(\hat{Q})\ge \sum_{i}P_{i}~C_{1}(\hat{Q}^{(i)}). \label{part1}\end{align}
To complete our proof we seek to show that 
\begin{align} 
 \sum_{i}P_{i}~C_{1}(\hat{Q}^{(i)})\ge C_{1}(\hat{X}).\end{align}
The explicit formulas for $C_{1}(\hat{Q}^{(i)})$ and $C_{1}(\hat{X})$, in terms of the elements of the $\hat{Q}^{(i)}$'s, are
 \begin{align} 
C_{1}(\hat{Q}^{(i)})&=2(|Q^{(i)}_{14}|-\sqrt{Q^{(i)}_{22}Q^{(i)}_{33}}),\\
C(\hat{X})&=2(|Q_{14}|-\sqrt{Q_{22}Q_{33}}),
\end{align}
where, for example, $Q_{14}=\sum_{i }P_{i} Q^{(i)}_{14}$. The following inequalities can be proved easily using the triangle and Cauchy-Schwarz inequalities:
\begin{align}
\sum_{i}P_{i}|Q^{(i)}_{14}|&\ge|\sum_{i}P_{i}Q^{(i)}_{14}|=|Q_{14}|,\\ \nn
\sum_{i}P_{i}\sqrt{Q^{(i)}_{22}Q^{(i)}_{33}}&\le\sqrt{\sum_{i}P_{i}Q^{(i)}_{22}} \sqrt{\sum_{j}P_{j}Q^{(j)}_{33}}\\ \nn
&=\sqrt{Q_{22}Q_{33}}~.
\end{align}
By subtracting the inequalities above one can show that 
\begin{align}
\sum_{i}P_{i}~C_{1}(\hat{Q}^{(i)})&\ge C_{1}(\hat{X}).\label{part2}
\end{align}
Two equations (\ref{part1}) and (\ref{part2}) complete our proof:
 \begin{align}C(\hat{Q})\ge C_{1}(\hat{X}).\end{align} 
Similarly one can show that $C(\hat{Q})\ge C_{2}(\hat{X})$ and therefore 
\begin{align} C(\hat{Q})\ge \text{Max}\{0,C_{1}(\hat{X}),C_{2}(\hat{X})\}=C(\hat{X})\label{finalresult},\end{align}
which is the inequality that we sought to prove. That is, we have proved that if one ignores the O-elements of a density matrix, the concurrence of the remaining X matrix is always smaller than or equal to the concurrence of the original density matrix.  \\

This inequality provides a sufficient condition that, if met, tells us that  the state is entangled and gives a lower bound on its entanglement. For a density matrix that takes the X form, the inequality in Eq.(\ref{finalresult}) becomes an equality. However, the converse in not necessarily true, i.e., it is possible that the matrix $\hat{Q}$ does not have the X form in the basis in which it is presented, and  yet the concurrence of the corresponding $\hat{X}$ is the same as the original density matrix  $\hat{Q}$. An example is given by
\begin{align}
\ket{\chi}=\frac{1}{2}\ket{\up,\up}+\frac{1}{\sqrt{3}} \ket{\up,\dn}+\frac{1}{\sqrt{6}} \ket{\dn,\up}+\frac{1}{2}\ket{\dn,\dn}.
\end{align}

An immediate question that comes up regards the possibility of a generalization of the current result  to the case beyond two-qubit density matrices. There are different ways to generalize concurrence to higher dimensions \cite{PhysRevA.62.032307,PhysRevA.64.042315,highdimcon}. We choose to work with the $\mathit{I~concurrence}$ that was first proposed by Rungta, et al. \cite{PhysRevA.64.042315} by generalizing spin flip super operator and it was later shown to be an entanglement monotone \cite{PhysRevA.67.012307}. From now on we simply refer to $\mathit{I~concurrence}$ as concurrence as it is widely referred to in the literature. If we refer to the two parties of the system as $A$ and $B$, the concurrence of a bipartite pure state is given by $\sqrt{2(1-Tr[\hat{Q}_{A}^{2}])}$ where 
\begin{align}
Q_{A}=Tr_{B}[Q]
\end{align}
 denotes the reduced density matrix of one of the subsystems. An arbitrary bipartite pure state can be written as 
\begin{align}
\ket{\Psi}=\sum_{i,k}a_{ik}\ket{i,k}.
\end{align}
The formula for the concurrence then can be written as follows \cite{highdimcon, PhysRevA.84.062322}:
\begin{align}
C(\ket{\Psi}\bra{\Psi})=2\sqrt{\sum_{i<j,k<l}|a_{ik}a_{jl}-a_{il}a_{jk}|^{2}}\label{conx}.
\end{align}
For mixed states the concurrence is defined by convex roof, as in Eq.(\ref{eqnMixedcon}). In analogy to the two-qubit case, we define 
\begin{align}
C_{ik,jl}(\hat{Q})=2\left(|Q_{ik,jl}|-\sqrt{Q_{il,il}Q_{jk,jk}}\right)
\end{align}
where, e.g., $Q_{ik,jl}=\bra{i,k}\hat{Q}\ket{j,l}$. For a pure state this quantity reduces to 
\begin{align}
C_{ik,jl}(\hat{Q})=2\left(|a_{ik}a_{jl}|-|a_{il}a_{jk}|\right),
\end{align}
which is always smaller than the concurrence defined in Eq.(\ref{conx}). $C_{ik,jl}(\hat{Q})$ then plays the same role that previously $C_{1}(\hat{Q})$ played in the proof for the two qubit density matrices and similarly one can show that

\begin{align}C(\hat{Q})\ge \text{Max}\{0, C_{ik,jl}(\hat{Q})~~ (i<j,k<l)\}.\end{align}
This is a sufficient condition for non-separability and the maximum of the $C_{ik,jl}$'s gives a lower bound for concurrence. \\

\begin{figure}[t]
 \vspace{0.25cm}
\includegraphics[width=\columnwidth]{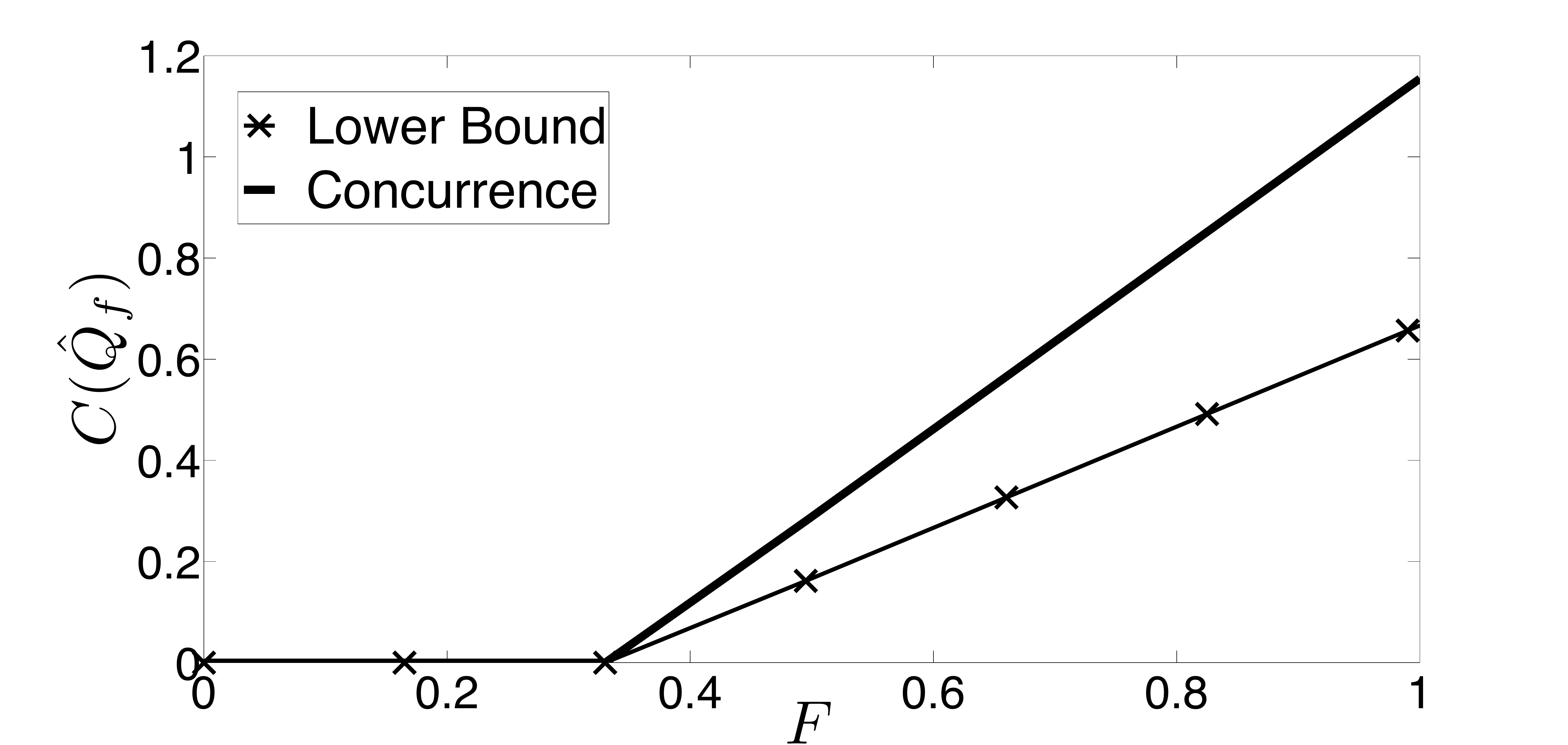}
\caption{The comparison between the actual concurrence and the lower bound for a $3\times 3$ isotropic state.}
\label{graph1}
\end{figure}

Let us also compare this lower bound with an exact formula. As mentioned, there is no algebraic formula (in a closed form) to calculate the concurrence of a bipartite mixed state of arbitrary dimension. However, for some special classes of mixed states the concurrence can be found exactly.  One class of such states is the class of isotropic states \cite{PhysRevA.59.4206,PhysRevA.64.062307}. This is a class of $U\otimes U^{*}$ invariant mixed states in $d\times d$ systems, where $U$ is a unitary operation and $U^{*}$ is its complex conjugate. Isotropic states take the form
\begin{align}
\hat{Q}_{f}=\frac{1-F}{d^{2}-1}\left(I-\ket{\psi_{+}}\bra{\psi_{+}}\right)+F\ket{\psi_{+}}\bra{\psi_{+}},
\end{align}
where $\ket{\psi_{+}}=\sqrt{1/d}\sum_{i}\ket{i,i}$ and $0\le F\le 1$. We plot the lower bound and the exact concurrence for $d=3$ in~Fig.~\ref{graph1}.  The concurrence of these states is known to be $\sqrt{2d/(d-1)}(F-1/d)$ and for $F\le1/d$ the isotropic states are separable  \cite{PhysRevA.67.012307}. Our lower bound for these states reads
\begin{align}
C(\hat{Q}_{f})\ge \frac{2}{d-1}(F-1/d) ~~~~~~ F\ge 1/d.
\end{align}
For $F\le1/d$, our lower bound gives zero.\\

In this paper, we established a lower bound on the concurrence of two qubits. The bound, unlike concurrence itself, does depend on the basis in which the state is given. Calculating this lower bound does not require any analytically challenging procedure, e.g., writing the state in a ``magic basis'' or even performing a diagonalization. This simplicity makes the defined lower bound an attractive tool in the analytical studies of entanglement dynamics with density matrices having no trivially zero elements. It may also prove to be a tool to confirm the entanglement of a state in an experimental setup with out performing a complete tomography and finding all the elements of the density matrix. 

Finally, our result brings up more questions that are left open for further investigation. One is the question regarding the cases when the equality holds. Of course if the density matrix, $\hat{Q}$, is in X form then Eq. (\ref{finalresult}) becomes an equality, the question is if the equality holds does it mean that there is a basis in which  $\hat{Q}$ becomes an X matrix? \\

The authors acknowledge discussions with C. Broadbent and J. H. Eberly. We acknowledge partial financial support from ARO W911NF-09-1-0385 and NSF PHY-0855701. \\

\bibliographystyle{apsrev4-1}
\bibliography{mybib}
\end{document}